\documentclass[conference]{IEEEtran}

\ifCLASSINFOpdf
  \usepackage[pdftex]{graphicx}
  \graphicspath{{Figures/}}
  \DeclareGraphicsExtensions{.pdf,.jpeg,.png}
\else
\fi

\usepackage{url}
\usepackage{tikz}
\usepackage{comment}
\usepackage{cite}

\usepackage{booktabs}

\usepackage{longtable}

\usepackage{makecell}
\usepackage{lscape}
\usepackage{afterpage}

\usepackage{tabularx}
\usepackage{multirow}

\usepackage{enumitem}
\setitemize{noitemsep,topsep=1pt,parsep=0pt,partopsep=0pt}

\begin{document}

\title{Before the Vicious Cycle Starts:\\Preventing Burnout Across SOC Roles\\Through Flow-Aligned Design}

\author{\IEEEauthorblockN{Kashyap Thimmaraju\IEEEauthorrefmark{1},
Duc Anh Hoang\IEEEauthorrefmark{1},
Souradip Nath\IEEEauthorrefmark{2},
Jaron Mink\IEEEauthorrefmark{2} and
Gail-Joon Ahn\IEEEauthorrefmark{2}}
\IEEEauthorblockA{\IEEEauthorrefmark{1}Technische Universität Berlin\\
Email: kashyap.thimmaraju@tu-berlin.de, duc-anh.hoang@campus.tu-berlin.de}
\IEEEauthorblockA{\IEEEauthorrefmark{2}Arizona State University\\
Email: \{snath8, jaron.mink, gahn\}@asu.edu}}

\IEEEoverridecommandlockouts
\makeatletter\def\@IEEEpubidpullup{6.5\baselineskip}\makeatother
\IEEEpubid{\parbox{\columnwidth}{
		{\fontsize{7}{7}\selectfont Workshop on Security Operation Center Operations and Construction (WOSOC) 2026 \\
			23 February 2026, San Diego, CA, USA \\
			ISBN 978-1-970672-03-9\\
			https://dx.doi.org/10.14722/wosoc.2026.23009\\
			www.ndss-symposium.org}
}
\hspace{\columnsep}\makebox[\columnwidth]{}}

\maketitle

\begin{abstract}
The sustainability of Security Operations Centers depends on their people, yet 71\% of practitioners report burnout and 24\% plan to exit cybersecurity entirely. Flow theory offers a lens for understanding this human factor challenge: when job demands misalign with practitioner capabilities—whether through excessive complexity or insufficient challenge—work becomes overwhelming or tedious rather than engaging. We argue that achieving this balance begins at hiring, the earliest intervention point in a practitioner's organizational journey. If job descriptions inaccurately portray role requirements, organizations risk recruiting underskilled practitioners who face chronic anxiety or overskilled ones who experience boredom. Both misalignments trigger burnout pathways, yet we lack empirical understanding of what skills and experience levels current SOC job descriptions actually specify, making it impossible to assess whether stated requirements set practitioners up for flow or frustration.

We address this gap by analyzing SOC job descriptions to establish the baseline of what challenge-skill profiles organizations claim to require. 
We collected and analyzed 106 public SOC job postings from November to December 2024 across 35 organizations in 11 countries, covering a range of SOC roles: Analysts, Incident Responders, Threat Hunters, and SOC Managers. 
Using Inductive Content Analysis, we coded certifications, technical skills, soft skills, tasks, and experience requirements (see Table~\ref{tab:skill-coverage} for an overview).
Our preliminary analysis revealed three key patterns: (1) Communication skills dominate requirements (50.9\% of 106 postings), substantially exceeding technical specifications like SIEM tools (18.9\% of 106) or programming (30.2\% of 106) suggesting that organizations prioritize communication and collaboration over purely technical capabilities. (2) Certification expectations are varied: CISSP leads (22\% of 106), but 43 distinct credentials appear with no universal standard, creating uncertainty for practitioners about which certifications merit investment. (3) Technical requirements show clear patterns: Python dominates programming (27\% of 106), Splunk leads SIEM platforms (14\% of 106), and ISO 27001 (13\% of 106) and NIST (10\% of 106) are the most cited standards, indicating an emerging consensus on core technical competencies that can guide both hiring decisions and training priorities.

This work represents the first stage of a research agenda to prevent burnout through sustained alignment of challenge-skill. The findings from this study establish an empirical baseline for what organizations claim to need, enabling validation studies that compare the stated requirements with actual practice. 
\end{abstract}

\IEEEpeerreviewmaketitle

\section{Introduction}
\label{sec:introduction}

\begin{table}[t]
\centering
\caption{Percentage of job descriptions (N=106) mentioning a subset of skill categories.}
\label{tab:skill-coverage}
\begin{tabular}{lrr}
\toprule
\textbf{Skill Category} & \textbf{Count} & \textbf{\% of Postings} \\
\midrule
Professional Skills & 84 & 79.2\% \\
Threat Intelligence & 58 & 54.7\% \\
Certifications & 36 & 34.0\% \\
Programming & 32 & 30.2\% \\
Security Standards & 22 & 20.8\% \\
SIEM Tools & 20 & 18.9\% \\
\bottomrule
\end{tabular}
\end{table}

Security Operations Centers (SOCs) are responsible for continuous monitoring, threat detection, incident response, and vulnerability management within organizations~\cite{rodman_soc_2024}. SOC practitioners work in environments where cognitive demands are high, decisions must be made rapidly under uncertainty, and the consequences of errors can be severe. In principle, such work should foster conditions for optimal human performance according to flow theory~\cite{flow-first-study, kotler-hbr, FRC-flow-def}: clear goals, immediate feedback, and meaningful challenges that develop practitioners' capabilities. Yet reality reveals a stark mismatch between this potential and actual outcomes.

Recent industry surveys document troubling patterns. Tines' 2023 Voice of the SOC report found that 71\% of SOC analysts experience burnout, with 64\% considering leaving their jobs within a year~\cite{tines2023voice}. A 2023 Devo/Wakefield Research survey revealed that 83\% of IT security professionals admit they or someone in their department has made errors due to burnout that led to security breaches, with 85\% anticipating they will leave their role due to burnout and 24\% planning to exit cybersecurity entirely~\cite{devo2023burnout}. For SOC leaders, 62\% of CISOs globally reported experiencing burnout in 2023~\cite{proofpoint2024voice}. The organizational consequences are severe: 23\% of SOC leaders report losing up to 19\% of their staff annually, with some organizations losing 40\% or more of their teams~\cite{sans2023breakingcycle}, and the average time to fill a SOC position is seven months, with 15\% of organizations reporting it takes two years or longer~\cite{sans2023breakingcycle}.

This persistent attrition and burnout represents not merely a human resources challenge but a fundamental threat to organizational security. When practitioners burn out and leave, institutional knowledge departs with them, leaving organizations vulnerable during extended hiring periods~\cite{human_performance_in_cybersecurity_protecting_2025, human_performance_in_cybersecurity_cybersecuritys_2025}. The human cost is equally concerning: exhaustion, cynicism, and a sense of inefficacy pervade the field~\cite{reeves-burnout-23,nepalBurnoutCybersecurityIncident2024, dupont_burnout_2025, shelton_state_2023, steve_shelton_state_2025}

Multiple theoretical frameworks converge on a common explanation: person-role misfit, particularly skill-challenge mismatch. Maslach's Areas of Worklife model identifies six domains that, when misaligned, trigger burnout: workload, control, reward, community, fairness, and values~\cite{leiterAreasWorklifeStructured2004}. Nepal et al.'s recent mixed-methods study of 35 incident responders identified workload, changing priorities, lack of clear goals and feedback, and inadequate recognition as key burnout factors
Sundaramurthy et al.'s ``Vicious Cycle Theory'' traces burnout in the SOC to its origins: analysts whose skills are underdeveloped are not empowered to develop those skills due to lack of managerial trust, creating a self-reinforcing cycle of underperformance and disengagement that ultimately leads to departure~\cite{sundaramurthy-soups}.

\textbf{Flow theory}, as articulated by Csíkszentmihályi, provides a complementary lens for understanding these dynamics~\cite{flow-first-study,flow-2013-study}. Flow---the state of optimal experience where individuals are fully immersed and performing optimally---requires a delicate balance between task challenge and individual skill. When challenges significantly exceed skills, anxiety results. When skills significantly exceed challenges, boredom sets in. Research on software developers has demonstrated that practitioners who are in the flow-state, experience both higher productivity and greater well-being~\cite{nodaDevExWhatActually2023,forsgrenDevExAction2024}. If practitioners are in roles or enter roles either underskilled or overskilled for the demands they face, the mismatch might prevent flow and establishes conditions for the burnout cycle Sundaramurthy et al.~identified.

\textbf{Identifying the intervention} point for preventing these cycles requires considering when different burnout factors emerge. Many factors manifest post-hiring as identified by Nepal et al.~\cite{nepalBurnoutCybersecurityIncident2024} in their study of incident responders who were past their ``honeymoon'' phase in their job: workload management, team dynamics, organizational support structures, and resource allocation practices develop over time within organizations. However, one critical factor can be addressed pre-hiring: the match between practitioner skills and role demands. If candidates enter SOC roles with skills appropriately calibrated to the challenges they will face, we establish the foundation for flow rather than frustration.
This requires clarity about what skills different SOC roles actually demand. Yet organizations may neither accurately communicate these requirements nor fully understand them themselves.

\textbf{Job descriptions (JDs)} currently serve as the primary mechanism through which organizations communicate role requirements to potential candidates. In principle, accurate JDs fulfill multiple functions: they help organizations attract appropriately skilled candidates, enable candidates to self-select into roles matching their capabilities, and establish clear expectations that support the ``clear goals'' dimension of flow theory~\cite{kotler-hbr}. However, we hypothesize that current JDs may not accurately reflect the skills and experience truly needed for SOC roles. Some may list requirements that are overly generic, others may emphasize certifications over practical competencies, and still others may fail to capture the resilience, communication and collaboration demands that research suggests are central to SOC effectiveness~\cite{vielberth2020security, reisser_security_2022}.

\textbf{Our vision} involves transforming the hiring pipeline to prevent burnout in SOCs. In particular, to validate and refine SOC job descriptions to support better challenge-skill alignment in hiring as a vicious cycle prevention strategy.
This requires a two-stage research approach: first, systematically analyzing what JDs currently demand; second, validating these stated requirements against actual role demands as understood by practitioners and hiring managers. Such (in)validation would identify mismatches: where JDs either under-specify or mis-specify requirements; and enable creation of more accurate JDs that support better hiring decisions. We hypothesize that accurate JDs, combined with flow-aligned hiring practices, can prevent the skill-challenge mismatch that triggers the Vicious Cycle~\cite{sundaramurthy-soups} and ultimately contributes to burnout. We note this hypothesis requires empirical validation through systematic research comparing stated requirements with actual practice and tracking outcomes over time.

\textbf{In this paper}, we take the critical first step: systematically analyzing what SOC job descriptions currently demand. We collected 106 public SOC job postings from November--December 2024 across Europe, Asia, and North America, spanning four key roles: Analysts, Incident Responders, Threat Hunters, and SOC Managers. Using an Inductive Content Analysis approach~\cite{vears_inductive_2022}, we coded certifications, technical skills, soft skills, tasks, and experience requirements. The coding process revealed challenges in source data \textit{and} analyzed data quality and consistency that necessitate careful interpretation. Nevertheless, this preliminary systematization provides observations about current JD practices and establishes an empirical foundation for subsequent validation studies.

We make the following \textit{contributions}:
\begin{itemize}
    \item Using an interdisciplinary approach we present a flow-based theory that explains why burnout occurs based on the vicious cycle theory for SOC burnout and how to prevent it from occurring in the SOC.
    \item We present a dataset of 106 SOC job descriptions from 11 countries across three continents, offering a broad empirical view of how organizations currently communicate \textbf{SOC role requirements}.
    \item We developed and publicly share a \textbf{preliminary codebook} for systematically analyzing SOC job descriptions, capturing certifications, technical skills, soft skills, tasks, and experience requirements across roles.
    \begin{center}
        \url{https://git.tu-berlin.de/wosoc-2026/soc-jd-analysis}
    \end{center}
    \item We document key observations about current JD practices, including communication emphasis, vagueness challenges, certification and tool patterns, experience requirement inconsistencies, and gaps in resilience-related attributes, establishing the empirical foundation for validation research and JD refinement efforts.
    \item We articulate a research agenda for validating JD requirements against actual practice and testing whether accurate JDs combined with flow-aligned hiring can prevent the skill-challenge-trust mismatch that contributes to SOC burnout.
\end{itemize}

\noindent\textbf{Ethical Considerations.}
All Job Descriptions were gathered from public sources and no people were involved in this research which exempted us from visiting the institutional review board. All the key concepts, ideas and data were created by us however, we leveraged AI-tools to improve the writing of this paper.

\noindent\textbf{Paper Structure.} The remainder of this paper is organized as follows. We provide background on the SOC organizational structure in Section~\ref{sec:background:soc-roles}, and on theoretical frameworks for understanding burnout in Section~\ref{sec:background:theory}. In Section~\ref{sec:background:cs-balance} we discuss the challenge-skill balance and its implications for SOC hiring. Our analysis of SOC job descriptions, including the data collection and coding methodology are detailed in Section~\ref{sec:jd-analysis}. Our preliminary findings are presented in Section~\ref{sec:preliminary-findings} followed by a discussion in Section~\ref{sec:discussion}. We then highlight related work in Section~\ref{sec:relatedwork} and then conclude in Section~\ref{sec:conclusion}.

\section{SOC Organizational Structure and Roles}
\label{sec:background:soc-roles}

An increasing number of organizations now operate a SOC team, where a specialized group is responsible for cybersecurity incidents. SOC teams are typically structured across four hierarchical tiers with distinct roles and their responsibilities, tasks, and skills~\cite{vielberth2020security, kokuluMatchedMismatchedSOCs2019}. Understanding these distinct roles is essential for analyzing how job descriptions communicate requirements and whether those requirements support appropriate challenge-skill matching.

\textbf{SOC Analysts} are responsible for the initial evaluation and prioritization of incoming security alerts, activities, and anomalies. They serve as the first line of defense and escalate relevant incidents to the next higher tier~\cite{vielberth2020security}. \textbf{Incident Responders} conduct in-depth investigations of the security incidents escalated by the SOC analyst. For these conducted analyses and assessments they require a wide variety of threat intelligence tools~\cite{hofbauer2024blue}. Additionally, they are tasked to develop and implement strategies as well as protocols, playbooks, to mitigate and recover from security incidents. \textbf{Threat Hunters} proactively track down previously undetected threats and develop advanced techniques and security tools. Furthermore, they handle escalated security incidents and perform in-depth system research to identify possible vulnerabilities and gaps to prevent security incidents from occurring. \textbf{SOC Managers} are tasked to lead the SOC team oversee processes, ensure governance and compliance with policies, and give strategic directions.

\section{Theoretical Frameworks for Understanding SOC Burnout}
\label{sec:background:theory}

Three complementary theoretical frameworks inform our understanding of burnout and attrition in SOC environments, each highlighting the critical role of person-role fit and skill-challenge alignment.

\noindent\textbf{Areas of Worklife.} Leiter and Maslach~\cite{leiterAreasWorklifeStructured2004} identify six domains that, when misaligned between person and work environment, contribute to burnout: workload, control, reward, community, fairness, and values. Among these, workload-capability mismatch is particularly relevant to hiring practices. Misalignment in workload occurs when job demands exceed the resources available to an individual~\cite{bakker_job_2007}. When practitioners are hired into roles for which they lack sufficient skills, they face workload demands they cannot adequately manage, establishing conditions for chronic stress rather than sustainable engagement. Similarly, lack of control manifests when practitioners have insufficient authority or autonomy to perform their responsibilities effectively. When underskilled practitioners are not trusted with decision-making authority appropriate to their role, they experience reduced control, compounding the workload mismatch. Accurate job descriptions that clearly communicate required skill levels can help prevent these misalignments by enabling better person-role matching from the outset.

\noindent\textbf{Vicious Cycle Theory.} Sundaramurthy et al.~\cite{sundaramurthy-soups} developed the Vicious Cycle Theory through an ethnographic study of SOC analysts, identifying how burnout develops through a self-reinforcing pattern. The cycle begins when analysts whose skills are underdeveloped are not empowered to develop those skills due to lack of managerial trust. This creates underperformance, which further reduces trust and empowerment, eventually leading to departure. The theory identifies skill-challenge mismatch at entry as a critical trigger: if practitioners begin with skills inadequately calibrated to their role's demands or unable to grow their skills, they enter the cycle immediately. Conversely, appropriate skill-role matching at entry enables practitioners to build trust, develop capabilities, and avoid the cycle entirely. This suggests that hiring decisions informed by accurate job descriptions could serve as an intervention point to prevent the Vicious Cycle before it starts.

\begin{figure}[t!]
    \centering
    \begin{tikzpicture}[scale=0.65]
        \draw[thick,->] (0,0) -- (8,0) node[below] at (4,0) {Skills};
        \draw[thick,->] (0,0) -- (0,8) node[left, rotate=90] at (-0.45,5) {Challenge};
        
        \draw[thick] (0,1.5) -- (7.0,7.5);
        \draw[thick] (1.5,0) -- (7.5,5.5);
        
        \node[anchor=south, font=\normalsize] at (3,6.5) {Anxiety (underskilled)};
        \node[font=\Large\bfseries, rotate=45] at (4,3.5) {Flow Channel};
        \node[anchor=north, font=\normalsize] at (6.5,1.5) {Boredom (overskilled)};
    \end{tikzpicture}
    \caption{The Flow Channel: Challenge-Skill Balance. When skills match challenges, individuals enter the flow state. Skills below challenge level lead to anxiety (underskilled), while skills exceeding challenges result in boredom (overskilled).}
    \label{fig:flow-channel}
\end{figure}
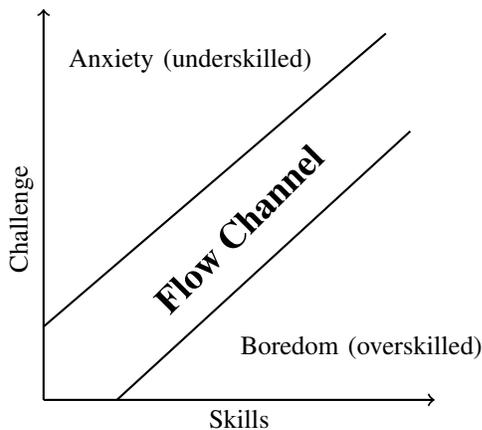

\noindent\textbf{Flow Theory.} Csíkszentmihályi~\cite{flow-first-study,flow-2013-study} describes flow as the psychology of optimal experience where individuals report being ``in the zone.'' Flow is an integral aspect of Positive Psychology and falls under the umbrella of Positive Engagement in the PERMA theory~\cite{perma-theory}. It is in the flow state where we feel and perform our best~\cite{FRC-flow-def} and is considered to be the antipode (opposite) of burnout~\cite{nepalBurnoutCybersecurityIncident2024}. Being in the flow state can be characterized by nine dimensions: challenge-skill balance, merging of action and awareness, clear goals, immediate feedback, complete concentration, sense of control, loss of self-consciousness, time transformation, and autotelic (``joy of doing'') experience. Among these, challenge-skill balance is foundational to entering the so-called flow channel illustrated in Figure~\ref{fig:flow-channel}. When task challenges slightly exceed one's skill level, flow occurs; when challenges significantly exceed skills, anxiety results; when skills significantly exceed challenges, boredom sets in. Research on software developers has demonstrated that practitioners who achieve flow experience both higher productivity and greater well-being~\cite{nodaDevExWhatActually2023,forsgrenDevExAction2024}.

\section{The Challenge-Skill Balance and Its Implications for SOC Hiring}
\label{sec:background:cs-balance}
The challenge-skill balance is particularly crucial for understanding work performance and well-being. 
While it is not always possible (or even unnecessary) to achieve flow in every task, persistent patterns of anxiety or boredom over extended periods can lead to burnout and attrition, potentially triggering the Vicious Cycle. When practitioners consistently operate in the anxiety zone, they may be unable to upskill effectively due to cognitive overload~\cite{cognitive-load-theory, vaishnav_impact_2025, COSS} and stress~\cite{thimmaraju_human_2025}. Alternatively, when practitioners persistently experience boredom, skill atrophy and disengagement may prevent them from taking on appropriately challenging work. In both cases, organizational factors such as inadequate training, lack of mentorship, or rigid role definitions may compound the challenge-skill mismatch.

This challenge-skill balance framework suggests important implications for SOC hiring. If practitioners enter roles with skills mismatched to the demands they face, they will struggle to achieve flow, instead experiencing either chronic anxiety (underskilled) or persistent boredom (overskilled), both very likely cases when analyzing SIEM alerts. Both states contribute to the burnout patterns documented in SOC environments. Establishing appropriate challenge-skill alignment from the outset requires clarity about knowing what JDs look like currently, what different SOC roles actually demand and whether organizations accurately communicate these requirements in their hiring materials. In the next section, we elaborate on our preliminary observations of 106 JDs from December 2024.

\section{SOC Job Description Analysis}
\label{sec:jd-analysis}

\subsection{Data Collection}
\label{sec:data-collection}
We collected 106 public SOC job postings between November and December 2024 from online job platforms (Indeed, StepStone, LinkedIn), company career pages and cybersecurity-specific recruiting companies. The dataset spans 35 organizations across Europe, Asia, and North America, including technology companies (Google, Microsoft, Meta), financial institutions (Citi), automotive manufacturers (Tesla, Mercedes-Benz), and specialized security firms (DCSO, Expel).

Job postings were included if they specified roles operating within or closely collaborating with SOCs. From initially screened advertisements, we selected most postings that articulated ``basic'' role requirements and responsibilities, this means we included some vague descriptions with insufficient detail. The final dataset tabulated in Table~\ref{tab:roledistribution} roughly comprises of 17 SOC Analyst positions, 38 Incident Responder positions, 39 Threat Hunter positions, and 12 SOC Manager positions. We note here that we made subjective decisions about classifying a JD into one of the four roles we consider as part of the SOC. A complete list of organizations and posting counts
appears in Table~\ref{tab:company-dataset} Appendix~\ref{appendix:companies} and also online.
\begin{center}
    \url{https://git.tu-berlin.de/wosoc-2026/soc-jd-analysis}
\end{center}
\begin{table}[t!]
\centering
\caption{Preliminary Role Distribution in Dataset (n=107)}
\label{tab:roledistribution}
\small
\begin{tabular}{@{}lc@{}}
\toprule
\textbf{SOC Role} & \textbf{n} \\
\midrule
SOC Analyst & 17 \\
Incident Responder & 38 \\
Threat Hunter & 39 \\
SOC Manager & 12 \\
\bottomrule
\end{tabular}
\end{table}

For each posting, we extracted role-relevant content including required qualifications, technical and soft skills, certifications, academic degrees, experience requirements, and responsibilities. Marketing content and benefit descriptions were excluded. We manually entered each posting into MAXQDA qualitative data analysis software, assigning each a unique identifier and standardizing the content into a consistent format for analysis.

\subsection{Coding Methodology}
\label{sec:coding-methodology}

Using Inductive Content Analysis~\cite{vears_inductive_2022} our coding process proceeded in four phases using MAXQDA.

\noindent\textbf{Phase 1: Open Coding.} Each job description was examined line-by-line to identify relevant keywords, allowing categories to emerge inductively from the data. While some categories appeared as explicit headings within postings (e.g., ``Required Skills''), additional patterns surfaced through systematic analysis. This initial iteration produced six main categories: Certifications, Degree, Experience, Soft Skills, Technical Skills, and Tasks.

\noindent\textbf{Phase 2: Category Refinement.} Technical Skills were subdivided into nine domains (Data Management, Forensics, Identity and Access Management, Incident Response, Programming and Frameworks, Security Management and Governance, Security Monitoring and Attack Detection, Technologies and Infrastructure, Threat Intelligence and Hunting). Tasks were organized into eight categories (Analysis and Evaluation, Communication and Collaboration, Development and Innovation, Documentation and Management, Implementation and Operations, Incident Response and Risk Management, Learning, Planning and Strategy). Experience was divided into years of experience and role-specific experience requirements.

\noindent\textbf{Phase 3: Automated Coding and Iteration.} Identified keywords were coded using RegEx patterns in MAXQDA, enabling consistent application across the full dataset. The process was iterated multiple times, combining manual corrections with automated procedures to capture keyword variations and resolve ambiguities, achieving theoretical saturation.

\noindent\textbf{Phase 4: Experience Categorization.} Several job descriptions did not specify experience explicitly in years, or mentioned experience with specific categories from Phase 2, which made coding experience very challenging, we consider improving this aspect as future work. The approach taken in the public codebook has been to develop experience bands that map to qualitative descriptors of ``experience'':
1--2 years (early career, entry level, bachelor's degree), 3--5 years (mid-level, moderate, several, significant, proficient, master's degree, PhD), 6--10 years (extensive, seasoned, advanced), and 11+ years (senior level, expert level). Postings with no experience requirements were coded as ``not required.'' Note that we did not analyze the experience category due to coding being unreliable.

\subsection{Data Quality Observations.}
This analysis faced several methodological constraints that affect interpretation. Job description quality varied substantially: some postings provided detailed competency frameworks with specific tools and technologies, while others used generic language such as ``strong analytical skills'' or ``excellent communication abilities'' without operational definitions.
Experience requirements presented particular challenges—many postings used qualitative descriptors (``proficient'', ``senior-level'', ``extensive experience'') that map ambiguously to actual capability levels, making systematic coding difficult. Coding was performed by a single researcher without inter-rater reliability testing, introducing potential interpretation variance and ambiguities. Most critically, stated requirements may not reflect actual hiring practices: recruiters may use job descriptions as aspirational specifications while accepting candidates meeting only subsets of requirements, or conversely, may prioritize unlisted characteristics during selection. These limitations highlight that \textit{our analysis captures what organizations \textit{claim} to require rather than what they \textit{actually} require or what skills predict success in these roles.}
The complete codebook and frequency distributions are available online.

\section{Our Preliminary Findings}
\label{sec:preliminary-findings}

The dataset reflects operational SOC workforce needs, with the majority of roles in hands-on security operations rather than management positions. Table~\ref{tab:skill-coverage} summarizes the overall coverage of different skill categories across the dataset. Due to coding quality, we do not present analyses for experience and technical skills but instead highlight some keywords.

\begin{table}[htbp]
\centering
\caption{Top 10 Professional Skills in SOC Job Descriptions (N=106). Emphasis column shows percentage of the 84 postings that mentioned professional skills.}
\label{tab:softskills_top10}
\begin{tabularx}{\columnwidth}{X r r r}
\toprule
Skill & Count & Cov. (\%) & Emph. (\%) \\
\midrule
Communication & 54 & 50.9 & 64.3 \\
Writing skills & 29 & 27.4 & 34.5 \\
Teamwork Abilities & 26 & 24.5 & 31.0 \\
Problem solving skills & 24 & 22.6 & 28.6 \\
Analytical skills \& mindset & 24 & 22.6 & 28.6 \\
Structured \& Reliable Work Style & 20 & 18.9 & 23.8 \\
Motivated & 15 & 14.2 & 17.9 \\
On Call Availability & 12 & 11.3 & 14.3 \\
Attention to Detail & 12 & 11.3 & 14.3 \\
Adaptable & 9 & 8.5 & 10.7 \\
\bottomrule
\end{tabularx}
\end{table}

\begin{table}[htbp]
\centering
\caption{Top 10 Professional Certifications in SOC Job Descriptions (N=106). Emphasis column shows percentage of the 36 postings that mentioned certifications.}
\label{tab:certs_top10}
\begin{tabularx}{\columnwidth}{X r r r}
\toprule
Skill & Count & Cov. (\%) & Emph. (\%) \\
\midrule
CISSP - Certified Information Systems Security Professional & 24 & 22.6 & 66.7 \\
CISM - Certified Information Security Manager & 13 & 12.3 & 36.1 \\
CEH - Certified Ethical Hacker & 11 & 10.4 & 30.6 \\
GCIH & 10 & 9.4 & 27.8 \\
Security+ & 10 & 9.4 & 27.8 \\
SANS & 9 & 8.5 & 25.0 \\
GIAC & 9 & 8.5 & 25.0 \\
OSCP - Offensive Security Certified Professional & 9 & 8.5 & 25.0 \\
CCNA & 7 & 6.6 & 19.4 \\
GCIA & 7 & 6.6 & 19.4 \\
\bottomrule
\end{tabularx}
\end{table}

\begin{table}[htbp]
\centering
\caption{Top 10 Programming Languages in SOC Job Descriptions (N=106). Emphasis column shows percentage of the 32 postings that mentioned programming languages.}
\label{tab:programming_top10}
\begin{tabularx}{\columnwidth}{X r r r}
\toprule
Skill & Count & Cov. (\%) & Emph. (\%) \\
\midrule
Python & 29 & 27.4 & 90.6 \\
Go & 10 & 9.4 & 31.2 \\
PowerShell & 10 & 9.4 & 31.2 \\
Java & 9 & 8.5 & 28.1 \\
C, C\#, C++ & 7 & 6.6 & 21.9 \\
BASH - Bourne Again Shell & 7 & 6.6 & 21.9 \\
Scala & 4 & 3.8 & 12.5 \\
PHP & 4 & 3.8 & 12.5 \\
JavaScript & 3 & 2.8 & 9.4 \\
Perl & 3 & 2.8 & 9.4 \\
\bottomrule
\end{tabularx}
\end{table}

\begin{table}[htbp]
\centering
\caption{SIEM Platforms in SOC Job Descriptions (N=106). Emphasis column shows percentage of the 20 postings that mentioned SIEM platforms.}
\label{tab:siem_all}
\begin{tabularx}{\columnwidth}{X r r r}
\toprule
Skill & Count & Cov. (\%) & Emph. (\%) \\
\midrule
Splunk & 15 & 14.2 & 75.0 \\
Microsoft Sentinel & 5 & 4.7 & 25.0 \\
Elasticsearch & 4 & 3.8 & 20.0 \\
Microsoft Defender & 4 & 3.8 & 20.0 \\
ArcSight & 3 & 2.8 & 15.0 \\
QRadar & 3 & 2.8 & 15.0 \\
Kibana & 1 & 0.9 & 5.0 \\
Google Chronicle & 1 & 0.9 & 5.0 \\
\bottomrule
\end{tabularx}
\end{table}

\begin{table}[htbp]
\centering
\caption{Top 10 Security Standards and Frameworks in SOC Job Descriptions (N=106). Emphasis column shows percentage of the 22 postings that mentioned security standards.}
\label{tab:secstandards_top10}
\begin{tabularx}{\columnwidth}{X r r r}
\toprule
Skill & Count & Cov. (\%) & Emph. (\%) \\
\midrule
ISO 27001 & 14 & 13.2 & 63.6 \\
NIST - National Institute of Standards and Technology & 11 & 10.4 & 50.0 \\
GDPR & 6 & 5.7 & 27.3 \\
OWASP & 4 & 3.8 & 18.2 \\
SOC 2 & 3 & 2.8 & 13.6 \\
FedRAMP & 3 & 2.8 & 13.6 \\
CIS & 2 & 1.9 & 9.1 \\
STIG & 2 & 1.9 & 9.1 \\
IEC62443 & 2 & 1.9 & 9.1 \\
NIS2 & 1 & 0.9 & 4.5 \\
\bottomrule
\end{tabularx}
\end{table}

\noindent\textbf{(1) Communication skills dominate requirements.} Communication skills appeared in 54 of 106 postings (50.9\%), with writing explicitly emphasized in 29 postings (27.4\%), as shown in Table~\ref{tab:softskills_top10}. Combined, more than 60\% of job descriptions emphasize communication capabilities. This substantially exceeds technical specifications: SIEM tools appeared in only 18.9\% of postings (Table~\ref{tab:siem_all}), while programming requirements appeared in 30.2\% of postings. Among soft skills (Table~\ref{tab:softskills_top10}), teamwork abilities (24.5\%), problem-solving (22.6\%), and analytical thinking (22.6\%) constitute the cognitive and collaborative competencies organizations seek, but communication remains the single most consistently mentioned requirement across all position types.

\noindent\textbf{(2) Certification expectations are varied with no dominant standard.} While 34.0\% of postings (36 of 106) mentioned at least one certification, as shown in Table~\ref{tab:certs_top10}, CISSP led among those requiring credentials (appearing in 66.7\% of certification-mentioning postings, or 22.6\% of all postings overall). CISM (12.3\% of all postings) and CEH (10.4\%) followed at substantially lower rates. The presence of 43 distinct certifications across the dataset suggests organizations recognize multiple paths to demonstrating competency. This diversity may reflect the varied technical domains within SOC work (monitoring, incident response, threat hunting, forensics) and the absence of a single universally recognized SOC credential. Notably, however, two-thirds of all postings (66.0\%) made no mention of certifications, suggesting credentials may be preferred rather than strictly required.

\noindent\textbf{(3) Technical requirements show technology-specific patterns.} Due to coding quality, we restricted our analysis to programming languages, SIEM tools, standards and certifications. When organizations specified technical requirements, clear patterns emerged. Python dominated programming requirements (27.4\% of all postings, appearing in 90.6\% of the 32 programming-mentioning postings), as shown in Table~\ref{tab:programming_top10}. PowerShell (9.4\% of all postings), Go (9.4\%), and Java (8.5\%) trailed substantially. SIEM platform specifications were relatively uncommon (18.9\% of postings overall), but when mentioned, Splunk dominated (75.0\% of SIEM-mentioning postings, or 14.2\% of all postings), with Microsoft Sentinel (4.7\% of all postings), Elasticsearch (3.8\%), and Microsoft Defender (3.8\%) mentioned occasionally (Table~\ref{tab:siem_all}). Security standards knowledge appeared in 20.8\% of postings, with ISO 27001 (13.2\% of all postings) and NIST (10.4\%) most frequently cited (Table~\ref{tab:secstandards_top10}). Among technical skill categories, incident response dominated (33.0\% of all postings), followed by forensics (22.6\%), reflecting the operational priorities of SOC work.

\textbf{While this analysis establishes what organizations claim to need in job descriptions}, it cannot answer the more fundamental question: do candidates meeting these stated requirements actually achieve challenge-skill balance on the job? A candidate with Python skills and CISSP certification might still experience anxiety if assigned tasks beyond their capability level, or boredom if relegated to routine monitoring work that underutilizes their expertise. Job descriptions specify credentials and competencies but not the challenges those credentials must address, nor the organizational context in which practitioners will apply their skills.

\section{Discussion}
\label{sec:discussion}

\subsection{Connecting Findings to the Flow Channel}
Several findings support flow-enabling conditions. Communication emphasis (50.9\% of 106 postings) aligns with research showing that feedback loops and clear goals—both essential flow dimensions—emerge from effective collaboration~\cite{Walker01012010, heyne_investigation_2011}. However, excessive communication demands can fragment attention and disrupt flow states~\cite{forsgrenDevExAction2024, nodaDevExWhatActually2023}, creating tension between coordination needs and deep focus. Python dominance (27.4\% of 106) enables automation that can reduce cognitive load on routine tasks, freeing mental capacity for appropriately challenging analytical work—a prerequisite for entering flow.
Other findings reveal barriers to flow assessment and achievement. Certification ambiguity (43 distinct credentials with no dominant standard) creates uncertainty that undermines candidates' ability to accurately gauge their skill levels relative to role demands. Resilience received rare mention (8.5\% of 106 postings) despite research emphasis on its importance for SOC work~\cite{vielberth2020security, nepalBurnoutCybersecurityIncident2024, reisser_security_2022}. We acknowledge that explicitly mentioning resilience in job descriptions may be challenging; however, it should be assessed during interviews to evaluate candidates' capacity to maintain flow under operational stress. Learning agility appeared infrequently (6.6\% of 106) despite rapid threat evolution requiring continuous upskilling to prevent practitioners' skills from falling below evolving challenge levels, a trajectory toward the anxiety zone.
Most critically, job descriptions specify credentials and tools but provide limited guidance about challenge complexity: the actual difficulty, ambiguity, and cognitive demands of daily tasks. While JDs serve as initial guides rather than complete specifications, this gap highlights the need for interview processes that systematically assess challenge-skill alignment. Our findings thus establish the baseline skill landscape from JDs, pointing toward the next research stage: examining how SOC interviews can capture the challenge dimensions necessary for accurate flow assessment and preventing the vicious cycle of underskilling, disempowerment, and burnout.

\subsection{Connecting Findings to the Vicious Cycle Theory and Challenge-Skill Balance}
To prevent a SOC practitioner from entering the ``Low skills'' state of the Vicious Cycle and leading to burnout, the challenge-skill balance plays a central role. Sundaramurthy et al.'s ethnographic research identified a destructive pattern: practitioners hired without sufficient skills become overwhelmed by operational demands, leading managers to withhold autonomy and trust, which further limits skill development opportunities and ultimately drives burnout and attrition~\cite{sundaramurthy-soups}. Our job description findings reveal both opportunities and risks for breaking this cycle at the hiring stage.
Previous research has identified that people in cybersecurity commonly find meaning and purpose in the work they do, which supports entering the flow channel and brings intrinsic motivation~\cite{thimmaraju_human_2025}. However, intrinsic motivation alone cannot sustain engagement when practitioners lack the skills to meet role demands or when skill growth stagnates. Our analysis shows that organizations specify extensive technical skill requirements—monitoring infrastructure (47.2\% of 106 postings), threat intelligence (54.7\% of 106), programming capabilities (27.4\% of 106 for Python): providing a foundation for initial skill-challenge alignment. Yet the absence of explicit learning expectations (learning agility mentioned in only 6.6\% of 106 postings) signals a potential gap in supporting continuous skill development necessary to prevent practitioners from falling behind evolving threat landscapes.
When candidates and organizations accurately assess initial fit for SOC roles, practitioners enter roles neither severely underskilled (risking immediate anxiety and loss of manager trust) nor severely overskilled (risking immediate boredom). However, maintaining this balance over time requires what Leiter et al.~terms ``control'', the autonomy to make decisions, apply judgment, and develop expertise through meaningful work~\cite{leiterAreasWorklifeStructured2004}. Communication emphasis in 50.9\% of 106 postings suggests collaborative environments where practitioners might exercise such autonomy, but job descriptions rarely make explicit the decision-making authority or growth opportunities that enable skill development, hopefully this is assessed during the interview.
Hence, preventing the Vicious Cycle requires managers to trust employees with appropriate autonomy once initial skill-challenge alignment is established through hiring. As practitioners develop technical capabilities in areas like SIEM platforms, scripting, or threat analysis, teams and organizations should consider progressively complex tasks that maintain flow. Without this progression, even well-matched hires eventually become overskilled for their assigned responsibilities, leading to disengagement: a different path to burnout than the underskilled trajectory, but equally damaging to retention and performance.

\subsection{Follow-Up Validation Studies Needed}
Our job description analysis reveals what organizations claim to need, but validation against actual practice remains essential. Do candidates meeting these stated requirements achieve challenge-skill balance? Does communication emphasis in job descriptions translate to actual collaborative work? Do Python skills predict success, or do other factors dominate? These questions require practitioner and manager surveys, longitudinal tracking of hiring outcomes, and systematic assessment of challenge-skill alignment in operational contexts. 

\subsection{Recommendations and Future Work}
We recommend, organizations to reduce credential ambiguity and clarify learning expectations in their JDs. Candidates should probe for challenge calibration and autonomy during interviews. 
Researchers should validate stated requirements against actual practice, longitudinal tracking of hiring cohorts, analysis of AI's impact on SOC skills, correlate job description accuracy with hiring outcomes and retention, and examine how AI integration affects role requirements.

\section{Related Work}
\label{sec:relatedwork}

Research on SOC roles, structure and skills have been conducted by a handful of researchers in the past, however, none to the best of our knowledge leverage their research to prevent burnout. In the following we highlight how our findings correlate with related work.

\subsection{SOC Roles and Structure}
Understanding SOC organizational structure and role definitions provides essential context for analyzing job requirements. Vielberth et al.~\cite{vielberth2020security} offer a comprehensive view of SOCs, highlighting key roles, required skills, and operational challenges from research in 2020. Building on this foundation, Hofbauer and Mayer~\cite{hofbauer2024blue} provide a structured overview of key SOC roles and tools through a systematic literature review with expert interviews. In particular, the authors enumerated several technical, management, and consulting SOC roles that inform our understanding of the SOC landscape. Complementing these taxonomic efforts, Reisser et al.~\cite{reisser_security_2022} employed an interview-based data collection methodology with eight practitioners, which contrasts with our more passive JD analysis. The authors observed that ``ability to work in a team'' was a frequently cited soft skill, which we also found in 30\% of the 106 JDs in our dataset. In general, our observations on SOC-related skills overlapped with these prior characterizations of the field.

\subsection{SOC Skills Assessment}
While role definitions establish what SOC positions exist, understanding how to assess practitioner capabilities remains a distinct challenge. From a high level view, our analysis of JDs correlates with findings by other researchers, particularly regarding communication as a key soft skill and CISSP as the most commonly cited certification. 

Radu et al.~\cite{radu_test_2025} designed and developed a set of tests to evaluate the skills of junior SOC analysts. We find that the approach of Radu et al.~to tailor the test of skills in the hiring process to better reflect on-the-job situations aligns with our flow-based challenge-skills balance concept.

Extending beyond SOC-specific roles, Sumner et al.~\cite{sumner_what_2023} presented a survey of the technical and professional skills needed for cybersecurity roles more broadly. As observed in our dataset and in Sumner et al., CISSP was the most-frequently mentioned certification. Similarly, both datasets confirm the presence of cybersecurity frameworks or standards, e.g., NIST and ISO. The concept of in-demand skills mentioned in the paper provides a useful way of prioritizing the availability of skills and what companies really want.

Ullah et al.~\cite{ullah_what_2025} conducted a large-scale analysis of JDs and Stack Overflow threads and, similar to us, also observed CISSP to be the most cited certification. However, with respect to programming languages, we observed Python more frequently compared to Java. Communication and project management were mentioned as the most important soft skills by Ullah et al., whereas our analysis was not coded with project management. We note that their dataset overlaps a noticeable amount with SOC-related JDs, suggesting convergence in industry requirements across cybersecurity domains.

\subsection{Burnout and Attrition in SOC}
While prior work establishes what skills SOC roles require and how to assess them, understanding why practitioners leave these roles is equally critical. The intense workload~\cite{nepalBurnoutCybersecurityIncident2024}, constant alert fatigue~\cite{hassan_nodoze_2019, alahmadi_99_nodate}, and high-pressure nature of SOC work~\cite{sundaramurthy-soups, nobles-burnout, nepalBurnoutCybersecurityIncident2024} contribute to stress, burnout, and cognitive overload, ultimately affecting the security of the organizations they protect~\cite{WorkFromHomeCOVID19Trajectories, technology_83_nodate, kokuluMatchedMismatchedSOCs2019, reeves-burnout-23, life-inside-perimeter, blackhat-europe-promon, tines, mediaMentalHealthCyber2022, 2022DevoSOC, VMwareReportWarns2022, state-of-ransomware, isc2-cybersecurity-workforce, CynetReveals942023}.

A majority of the research thus far has identified the presence of burnout~\cite{sundaramurthy-soups, nepalBurnoutCybersecurityIncident2024, reeves-burnout-23, thimmaraju_human_2025, dupont_burnout_2025}. However, research is sparse on preventions tailored to SOCs. In this paper, we leverage an existing theory for burnout in the SOC~\cite{sundaramurthy-soups} and present a systematic approach to preventing burnout through flow-state theory, beginning with an empirical analysis of the job requirements that establish initial challenge-skill alignment.

\section{Conclusion}
\label{sec:conclusion}
We set out to address a fundamental question: why does SOC work, which should enable optimal human performance through clear goals, immediate feedback, and meaningful challenges, instead produce burnout rates exceeding 70\%? We hypothesized that person-role misfit, particularly skill-challenge mismatch, triggers the Vicious Cycle~\cite{sundaramurthy-soups}, leading practitioners from underskilling to disempowerment to lack of creativity to lack of growth resulting in burnout and attrition. If accurate job descriptions could support better challenge-skill alignment at hiring, we reasoned, organizations might prevent this cycle before it starts. However, testing this hypothesis required first understanding what SOC job descriptions currently communicate about role requirements.

Our preliminary analysis of 106 SOC job postings from November to December 2024 across 11 countries revealed three key patterns. First, communication skills dominate requirements, appearing in 50.9\% of 106 postings and substantially exceeding technical specifications such as SIEM tools (18.9\% of 106) or programming languages (30.2\% of 106). Second, certification expectations are high but varied, with CISSP appearing most frequently (22.6\% of 106 postings) among 43 distinct credentials, yet two-thirds of postings mention no certifications at all. Third, technical requirements show clear technology-specific patterns, with Python dominating programming needs (27.4\% of 106), Splunk leading SIEM platforms (14.2\% of 106), and ISO 27001 (13.2\% of 106) and NIST (10.4\% of 106) representing the most frequently cited security standards. These findings establish an empirical baseline for what organizations claim to need, though we acknowledge significant limitations: coding was performed by a single researcher without inter-rater reliability testing, experience and technical requirements proved particularly challenging to systematize due to vague qualitative descriptors, and stated requirements may diverge substantially from actual hiring practices.

The broader implications extend beyond hiring optimization to workforce sustainability and organizational security. If organizations can accurately communicate role requirements and assess challenge-skill fit during hiring, they establish conditions for practitioners to enter the flow channel rather than immediately experiencing anxiety or boredom. Over time, as skills develop, teams and organizations should progressively increase task complexity to maintain flow, preventing the skill stagnation that leads to disengagement or the overwhelming cognitive load that triggers the Vicious Cycle. This requires not just better job descriptions, but flow-aligned interview processes that assess challenge-skill readiness, adaptive role management systems that adjust responsibilities as practitioners develop, and potentially AI-assisted workflows that handle routine cognitive work while preserving appropriately challenging analytical tasks. Preventing SOC burnout thus demands systematic intervention across the employment lifecycle, starting with but extending far beyond hiring.

\section*{Acknowledgment}
The authors thank the anonymous reviewers for their feedback. First author K.~T.~thanks Sybe Izaak Rispens for his feedback on an earlier version of this paper, Jean-Pierre Seifert for his support, and acknowledges financial support by the German Federal Ministry of Research, Technology and Space in the framework of the Q-Fiber project number 16KISQ124.

\bibliographystyle{IEEEtran}
\bibliography{references,MagnumOpus}

\appendix
\label{appendix:companies}
\begin{table}[th]
\centering
\caption{Organizations and Job Posting Counts in Dataset (n=106)}
\label{tab:company-dataset}
\begin{tabular}{@{}lc@{\hspace{2em}}lc@{}}
\toprule
\textbf{Organization} & \textbf{n} & \textbf{Organization} & \textbf{n} \\
\midrule
360 SOC Inc & 1 & Mercedes-Benz & 2 \\
Asembia & 1 & Meta & 2 \\
Atruvia AG & 2 & Microsoft & 6 \\
Baker Hughes & 2 & MHP (Porsche) & 3 \\
Citi & 4 & Motorola Solutions & 2 \\
Cloudflare & 4 & NTT Data & 6 \\
Computacenter & 2 & Otto GmbH & 1 \\
DAZN & 2 & Profiler GmbH & 5 \\
DCSO & 4 & r-tec IT Security & 4 \\
Eclaro & 1 & SilverSky & 3 \\
Encora & 1 & Tesla & 4 \\
EthicalHat & 2 & Trustmi & 1 \\
Expel & 4 & thyssenkrupp AG & 3 \\
Google & 6 & UK Carl Gustav Carus & 3 \\
GSK & 1 & Wiz & 2 \\
ION & 2 & Wipro & 4 \\
LB IT Niedersachsen & 1 & X (formerly Twitter) & 5 \\
 &  & Unknown & 11 \\
\bottomrule
\multicolumn{4}{@{}l@{}}{\footnotesize UK = Universitätsklinikum; LB = Landesbetrieb} \\
\end{tabular}
\end{table}

\end{document}